\documentclass[a4paper,10pt,twocolumn]{revtex4}
\usepackage[english]{babel}
\usepackage{indentfirst}
\usepackage{graphicx}
\usepackage{amsmath}
\usepackage{amssymb}
\usepackage{amsthm}
\usepackage{mathrsfs}
\usepackage{bbold}
\usepackage{lscape}
\usepackage{bm}
\usepackage{bbm}
\usepackage{float}
\usepackage{longtable}
\usepackage{graphics}
\usepackage{color}
\usepackage[utf8x]{inputenc}

\begin{document}

\title{Quantum quench of Kondo correlations in optical absorption}


\author
{C. Latta$^{1\ast}$, F. Haupt$^1$, M. Hanl$^2$, A. Weichselbaum$^2$, M. Claassen$^1$, W. Wuester$^1$, P. Fallahi$^1$, S. Faelt$^1$, L. Glazman$^3$, J. von Delft$^2$, H. E. T\"ureci$^{4,1}$ A. Imamoglu$^{1\ast}$
\normalsize{$^{1}$ Institute of
  Quantum Electronics, ETH-Z\"urich, CH-8093 Z\"urich, Switzerland,}\\
\normalsize{$^2$ Arnold Sommerfeld Center for Theoretical Physics,}\\
\normalsize{Ludwig-Maximilians-Universit\"at M\"unchen, D-80333 M\"unchen,Germany,}\\
\normalsize{$^3$ Sloane Physics Laboratory, Yale
  University, New Haven, CT 06520, USA}\\
\normalsize{$^4$ Department of Electrical Engineering, Princeton University, Princeton, New Jersey 08544, USA}
\\
\normalsize{$^\ast$To whom correspondence should be addressed;} \\
\normalsize{E-mail:  clatta@phys.ethz.ch or imamoglu@phys.ethz.ch}
}


\date{}

\newcommand{\jvd}[1]{{\color{magenta}{#1}}}
\newcommand{\comment}[1]{{\color{blue}{\textbf{[Comment: #1]}}}}
\newcommand{\todo}[1]{{\color{red}{\textbf{[ToDo: #1]}}}}
\newcommand{\scrap}[1]{{\color{red}{#1}}}
\newcommand{\latta}[1]{{\color{magenta}{#1}}}
\newcommand{\atac}[1]{{\color{orange}{#1}}}
\newcommand{\Uee}{U_{\rm ee}}
\newcommand{\Ueh}{U_{\rm eh}}
\newcommand{\Tk}{T_{\rm K}}
\newcommand{\TFR}{T_{\rm FR}}



\baselineskip24pt \maketitle {\bf The interaction between a single
confined spin and the spins of a Fermionic reservoir leads to one of
the most spectacular phenomena of many body physics -- the Kondo
effect\cite{kondo1964,KouwenhovenG01}. Here we report the
observation of Kondo correlations in optical absorption measurements
on a single semiconductor quantum dot tunnel-coupled to a degenerate
electron gas. In stark contrast to transport
experiments\cite{goldhaber1998,CronenwettOK98,vanderwiel2000},
absorption of a single photon leads to an abrupt change in the
system Hamiltonian and a quantum quench of Kondo correlations. By
inferring the characteristic power law exponents from the
experimental absorption line-shapes, we find a unique signature of
the quench in the form of an Anderson orthogonality
catastrophe\cite{MahanBook,anderson1967}, originating from a
vanishing overlap between the initial and final many-body
wave-functions. We also show that the power-law exponents that
determine the degree of orthogonality can be tuned by applying an
external magnetic field which gradually turns the Kondo correlations
off\cite{tureci2010}. Our experiments demonstrate that optical
measurements on single artificial atoms offer new perspectives on
many-body phenomena previously studied exclusively using transport
spectroscopy. Moreover, they initiate a new paradigm for quantum
optics where many-body physics influences electric field and
intensity correlations.}

Optical spectroscopy of single quantum dots (QD) has demonstrated
its potential for applications in quantum
information processing, particularly in the realization of single
and entangled photon sources\cite{Michler00,dousse2010}, coherent
spin qubits\cite{press2008,kim2010} and a spin-photon
interface\cite{yilmaz2010,claassen2010}. Even though recent
experiments have established this system as a new paradigm for
solid-state quantum optics, all of the striking experimental
observations to date could be understood within the framework of
single- or few-particle physics enriched by perturbative coupling to
reservoirs involving either phonons, a degenerate electron gas
\cite{dalgarno2008,kleemans2010}, or nuclear spins
\cite{latta2009,xu2009}).

We present differential transmission (DT) experiments\cite{hogele04}
on a single charge-tunable QD that reveal optical signatures of the
Kondo effect\cite{kondo1964,KouwenhovenG01}. In contrast to prior
experiments\cite{Atature06,kleemans2010}, the tunnel coupling
between the QD and a nearby degenerate electron gas, which we refer
to as the Fermionic reservoir (FR), is engineered to be so strong
that the resulting exchange interactions cannot be treated within
the framework of a perturbative system-reservoir theory: In the
initial state, the ``system'' (QD spin) is maximally entangled with
the FR, forming a screened singlet.

The fundamentally new feature that differentiates the results we present from all
prior transport based investigation of the Kondo
effect\cite{goldhaber1998,CronenwettOK98,vanderwiel2000}, is the realization of a
quantum quench of the local Hamiltonian; in our experiments, photon absorption
abruptly turns the exchange interaction between the QD electron and the FR off,
leading to the destruction of the correlated QD-FR singlet that otherwise acts as
a local scattering potential for all FR electrons. As was shown by
Anderson\cite{MahanBook,anderson1967}, the overlap between $N$-electron FR states
with and without a local scattering potential scales as $N^{-\alpha}$ with $\alpha
> 0$. This reduced overlap, termed Anderson orthogonality catastrophe (AOC), leads
to a power-law tail in absorption if the scattering potential is turned on or off
by photon absorption. Here, we determine the AOC induced power-law exponents in
absorption line shape that uniquely characterize the (quench of) Kondo
correlations. Moreover, by tuning the applied laser frequency, we observe both the
perturbative and non-perturbative regimes of the Kondo effect in one absorption
line shape, without having to change the FR (electron) temperature $\TFR$.


\textbf{Experimental setup.} The schematic of the QD sample we study
is shown in Figure~1a: a gate voltage $\rm{V_g}$ applied between a
top Schottky gate and the degenerate electron gas allows us to tune
the charging state of the QD\cite{warburton2000}. Figure~1B shows
the photoluminescence (PL) spectrum as a function of $\rm{V_g}$,
where different discrete {\it charging plateaux} are clearly
observable. In this paper we focus on the X$^-$ plateau, for which
the QD is single-electron charged and the influence of the FR on the
QD PL dispersion and linewidth is strongest.  The $\rm{X}^{-}$
optical transition couples the {\it initial configuration},
containing on average one electron in the QD, to a {\it final
configuration}, containing on average two electrons and a
valence-band hole (a negatively charged trion). This transition can
be described within the framework of an excitonic Anderson model
(EAM) \cite{helmes2005,tureci2010}, depicted schematically in
Fig.~2C (and described explicitly in Supporting Online Material). It
is parameterized by the energy $\varepsilon$ of the QD electron
level w.r.t. to the Fermi level, the on-site Coulomb repulsion
$\Uee$, the tunnel rate $\Gamma$ between QD and FR, the
half-bandwidth $D$ of the FR, and the electron-hole Coulomb
attraction $\Ueh$. The latter is relevant only in the final
configuration, where it effectively lowers the electron level energy
to $\varepsilon - \Ueh$, thus ensuring the double occupancy of the
electron level in the final configuration. A Hartree-Fock estimate
from the PL data in Figure 1b yields $\Ueh \simeq \Uee + 4$~meV.

\textbf{Absorption line shape.}  The inset of Figure 2a shows high resolution
laser absorption spectroscopy on the same QD across the X$^-$ single electron
charging plateau (see Supporting Online Material). Here, we parameterize $V_g$ in
terms of $\varepsilon$, normalized and shifted such that $\varepsilon =
-\frac{1}{2} \Uee$ for the gate voltage where the absorption contrast is maximal.
Instead of the usual linear dc-Stark shift of the absorption peak that is
characteristic of charge-tunable QDs, we find a strongly non-linear
$\varepsilon$-dependent shift of the $\rm{X}^{-}$ transition
energy\cite{dalgarno2008,kleemans2010}, which measures the energy difference
between the final and initial ground states. This energy shift arises from a
renormalization of the initial state energy \cite{anderson1961} due to virtual
tunneling between the singly-occupied QD and FR (analogous to the Lamb shift of
atomic ground states). The final trion state energy, on the other hand, is hardly
affected by virtual tunneling processes, due to large $\Ueh - \Uee$. As depicted
in Figure 2c and explained in its caption, this renormalization-induced red-shift
of the initial state is strongest at the plateau edges and leads to an
$\varepsilon$-dependent blue-shift of the optical resonance frequency.  The latter
can be used to determine the EAM model parameters: $\Uee$=7.5 meV,
$\Gamma=0.7$~meV, and $D = 3.5$~meV. NRG calculations for the transition energy
(Figure~2a, solid blue line) give excellent agreement with the experimental data
(blue symbols).



Figure~3\textbf{a} shows on a log-log scale, the blue tail of the normalized
absorption line shapes, where the applied laser frequency is larger than the
transition energy, for the four values of gate voltage indicated by arrows in
inset of Figure~2a. Inset to Figure~3a compares the full unnormalized absorption
line shapes for the identical gate voltages in linear scale; the red absorption
tail allows us to determine the temperature of the FR as $T_{\rm FR} = 180$~mK$ =
15.6 \mu$eV (see Supporting Online Material). The strong variation of the peak
absorption strength in Figure~3a inset is a consequence of the exponential
dependence of the Kondo temperature $\Tk (\varepsilon) = \sqrt{\Gamma D} e^{-[1 -
(2\varepsilon/\Uee  + 1)^2](\pi \Uee/ 8 \Gamma )}$ on the gate voltage
$\varepsilon$. For this QD, $\Tk$ varies between $24\mu$eV and $464 \mu$eV
\footnote{$\Tk$ looses significance for the black curve, for which the QD-FR
system is in the mixed-valence regime.}. All line shapes carry the signatures of
an optical interference effect induced by the sample structure (causing some
lineshapes to become negative for small red detunings), and of independently
measured fluctuations in gate voltage; both effects have been taken into account
in the calculated line shapes (see Supporting Online Material). Calculating the
line shapes by NRG (solid lines) without any further fit parameters, we find
remarkable agreement with experiment for all four lineshapes depicted in
Fig.~3\textbf{a}, demonstrating the validity of the excitonic Anderson
Model\cite{tureci2010} for the coupled QD-FR system.

\textbf{Perturbative regime.} To obtain better insight into the underlying
physics, we first consider the absorption line shape for blue detunings satisfying
$\nu > {\rm max}(\TFR,\Tk)$, where a perturbative description is possible. All
four line shapes in Fig.~3a impressively show the $\nu^{-1}$ dependence expected
for the perturbative regime. The origin of the $\nu^{-1}$ tail can be understood
as arising from a two-step process, where a virtual excitation of a valence-band
electron to the conduction band is followed by excitation of an electron-hole pair
in the FR via an effective spin-exchange coupling. The energy of the electron-hole
pair is equal to $\nu$, ensuring energy conservation. The lowest-energy electron
in the FR, which can contribute to this process, has an energy $-\nu$ (we took the
Fermi energy to be $0$). Assuming a constant density of states around the Fermi
level of the FR, the effective bandwidth of electrons contributing to this
two-step absorption process at a detuning $\nu$ is equal to $\nu$. The
electron-hole pair generation therefore contributes an additional phase space
factor $\propto \nu$ to the Lorentzian broadened QD trion transition
\cite{tureci2010}, such that the overall lineshape has approximately a $1/\nu$
tail.

\textbf{Scaling collapse.} For gate voltages such that the initial ground state is
a Kondo singlet, the absorption line shape, normalized to its value at $\nu =
\Tk$,  is expected \cite{tureci2010} to be a universal function of $\nu/\Tk$ in
the regime $\TFR \ll \nu \ll \Uee$. To confirm this prediction, Fig.~3c shows
$A(\nu)/ A(\Tk(\varepsilon))$ as a function of $\nu/\Tk$ for three of the line
shapes of Fig.~3a (but omitting the black one, for which the coupled system is in
the mixed valence regime). A striking scaling collapse is evident
\footnote{Deviations from scaling for $\nu < \TFR$ are expected, but masked by an
insufficiently small signal-to-noise ratio of the experimental data.}. Multiplying
$\Tk$ by a constant factor $\xi$ for all three $\varepsilon$ values yields a
scaling collapse of nearly identical quality, provided $1 \le \xi \le 5$. This
reflects the fact that $T_{\rm K}$, being a crossover scale, is defined only up to
an arbitrary pre-factor of order one.

\textbf{Strong-coupling Kondo regime.} In the limit $\TFR < \nu < \Tk$, a
perturbative description of the lineshape is no longer valid. In the initial
configuration, the exchange interaction between QD and FR induces a {\it Kondo
screening cloud} that forms a singlet with the QD spin. This acts as a scattering
potential that induces \emph{strong} phase shifts for those low-energy fermionic
excitations whose energies are within $\Tk$ from the Fermi level. In the final
configuration after photon absorption, the QD has two electrons in a \emph{local}
singlet state. Therefore the Kondo screening cloud, and the scattering potential
for FR electrons constituted by it, disappear in the long time limit: the
corresponding ground state wave-function is a tensor product of the local singlet
and free electronic states, with only \emph{weak} phase shifts.  Since the initial
and final FR phase shifts differ (as depicted schematically in Fig.~2\textbf{d}),
the FR does not remain a spectator during the X$^-$ transition: instead, the
transition matrix element between the ground states of the initial and final
configurations is vanishingly small. This leads to an AOC which manifests itself
by transforming a delta-function resonance (of an uncoupled QD) into a power-law
singularity \cite{MahanBook} of the form $\nu^{-\eta}$, where the exponent $\eta$
characterizes the extent of AOC.  For $\TFR \ll \nu \ll \Tk$, the absorption
lineshape of the X$^-$ transition is expected to show an analogous power-law
singularity. The exponent $\eta$ is predicted \cite{helmes2005,tureci2010} to
range between 0 and 0.5 (assuming no magnetic field), with $\eta \simeq 0.5$ being
characteristic for a Kondo-correlated initial state and an uncorrelated final
state. This lineshape modification is a consequence of a redistribution of the
optical oscillator strength, associated with the fact that the FR wave-function in
the Kondo correlated initial state has finite overlap with a range of final states
consisting of electron-hole pair excitations out of a non-interacting FR.

If $\TFR \ll \Tk$ and the optical detuning is reduced below $\Tk$, the lineshape
is predicted to smoothly cross over from the perturbative $1/\nu$ tail to the
strong-coupling AOC $1/\nu^{0.5}$ power law just discussed. This crossover is
illustrated in Fig.~3\textbf{b} (dashed lines) by NRG calculations, performed at
$\TFR=0$ for the four $\varepsilon$-values of Fig.~3\textbf{a}: Remarkably,
despite drastic differences in the $\nu > \Tk$ tails due to different values of
$\Tk(\varepsilon)$, all four lineshapes show similar power-law exponents around
$\eta \simeq 0.5$ for $\nu \ll \Tk$, in accordance with predictions
\cite{tureci2010} based on Hopfield's rule (see Supplementary Information). For
nonzero temperature, however, the $1/\nu^{0.5}$ power law is cut off and saturates
once $\nu$ decreases past $\TFR$ (Fig.~3b, solid lines), because of thermal
averaging over initial states with excitation energies $\lesssim \TFR$.

\textbf{Magnetic field-tuning of Kondo correlations.} A direct extraction of the
$1/\nu^{0.5}$ power law from the measured data is difficult due to the small
accessible experimental window $\TFR < \nu < \Tk$. Nevertheless, we are able to
determine the power-law exponent corresponding to our data accurately by using the
fact that the detailed form of the line shape sensitively depends on the exponent
$\eta$ which can be tuned by an external magnetic field\cite{tureci2010}.
Figure~4a shows the measured absorption line shape of a second QD  at
$B_{ext}=0$~Tesla (black squares) and the line shapes of the blue (blue  dots) and
red (red triangles) Zeeman/trion transition at $B_{ext}=1$~Tesla in a linear
scale. We have determined the parameters of this QD to be $\Uee = 7.5$~meV,
$\Gamma = 1$~meV, $D = 6.5$~meV and $\Ueh = 3/2 \Uee$; the data were taken at
$\TFR = 15.6 \mu$eV for $\varepsilon/\Uee=-0.43$ where $T_K = 140 \mu$eV. While
the peak contrast at $B_{ext} = 1$Tesla for the blue (red) trion transition
increases (decreases) by a factor $\sim 2$, the area under the absorption curve
increases (decreases) by less than $20 \%$; this observation proves that the
change in the line shape is predominantly due to a line narrowing associated with
a increase (decrease) of the power-law exponent $\eta$. Conversely, the small
change in the area under the absorption curve despite having $B_{ext} = 2.5 \TFR$
demonstrates that the initial state is a Kondo singlet that suppresses the
electron Zeeman splitting. The inset to Fig.~4a shows the $B_{ext}$ dependence of
the peak absorption contrast for the blue and red trion transitions; while the
agrement with the predictions of NRG is excellent for $B_{ext} \le 1.5$Tesla, the
blue trion contrast exhibits oscillations for higher $B_{ext}$; this is most
likely a consequence of the modification of the FR density of states at high
fields in Faraday configuration. Fig.~4b shows the normalized line shape in a
log-log plot together with the results of the NRG calculation (solid lines):
$\eta$ values that we determine from these plots range from $0.31$ (red trion) to
$0.66$ (blue trion), and prove the sensitivity of the measured line shapes to the
AOC determined power-law exponents. These experiments unequivocally demonstrate
the magnetic tuning of the AOC exponent for the first time.

The remarkable agreement between our experimental data depicted in
Figs.2-4 and the NRG calculations clearly demonstrate Kondo
correlations between a QD electron and the electrons in a FR, as
predicted by the excitonic Anderson model. The optical probe of
these correlations unequivocally show the signatures of Anderson
orthogonality physics associated with the quantum quench of Kondo
correlations. Our experiments establish the potential of single
optically active QDs in investigating many-body physics. In
addition, they pave the way for a new class of quantum optics
experiments where the influence of simultaneous presence of
non-perturbative coherent cavity/laser coupling and Kondo
correlations on electric field and photon correlations could be
investigated.

\bibliography{scibib}

\bibliographystyle{Science}

\clearpage
\begin{figure}[t]
   \includegraphics[scale=1]{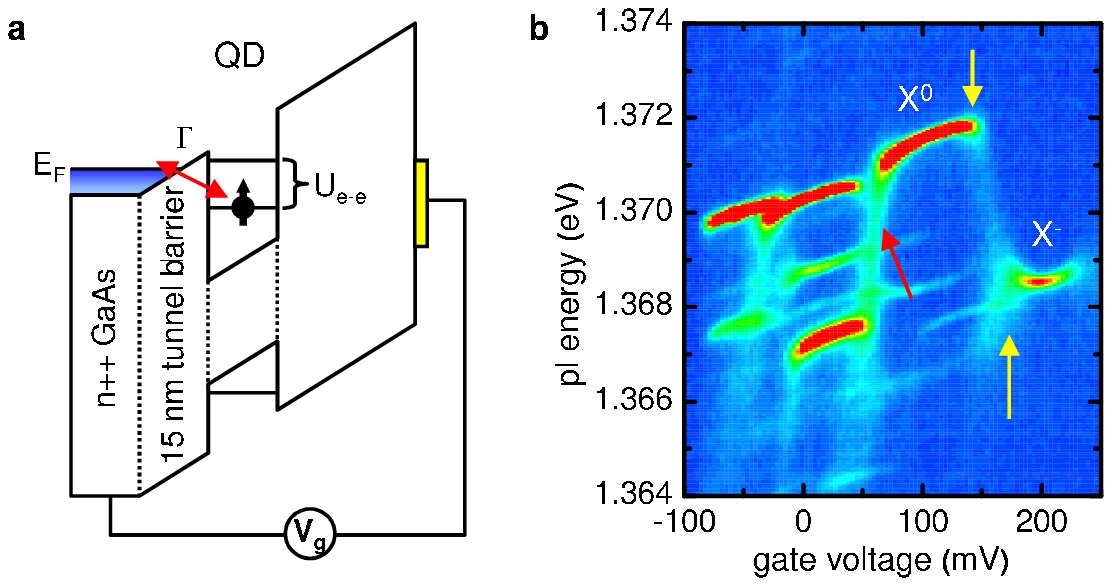}
    \caption{ A single quantum dot strongly coupled to a fermionic reservoir.
    (a) Band structure of the device. The QDs are separated by a 15
nm tunnel barrier from a $\rm{n}^{++}$-doped GaAs layer (Fermi sea).
A voltage $\rm{V_g}$ applied between the electron gas and a
semi-transparent NiCr gate on the sample surface controls the
relative value of the QD single-particle energy levels with respect
to the Fermi energy $\rm{E_F}$.
(b) Low temperature (4 K)
photoluminescence spectrum of a single QD as a function of
$\rm{V_g}$. The interaction of the QD electron with the Fermi sea
leads to a broadening of the photoluminescence lines at the plateau
edges (yellow arrows) and indirect recombinations of a QD hole and a
Fermi sea electron (red arrow). Indirect transitions are identified
by the stronger $\rm{V_g}$ dependence of the transition energy
compared to direct transitions.}
\end{figure}

\newpage

\newpage
\clearpage

\begin{figure}[t]
   \includegraphics[scale=1]{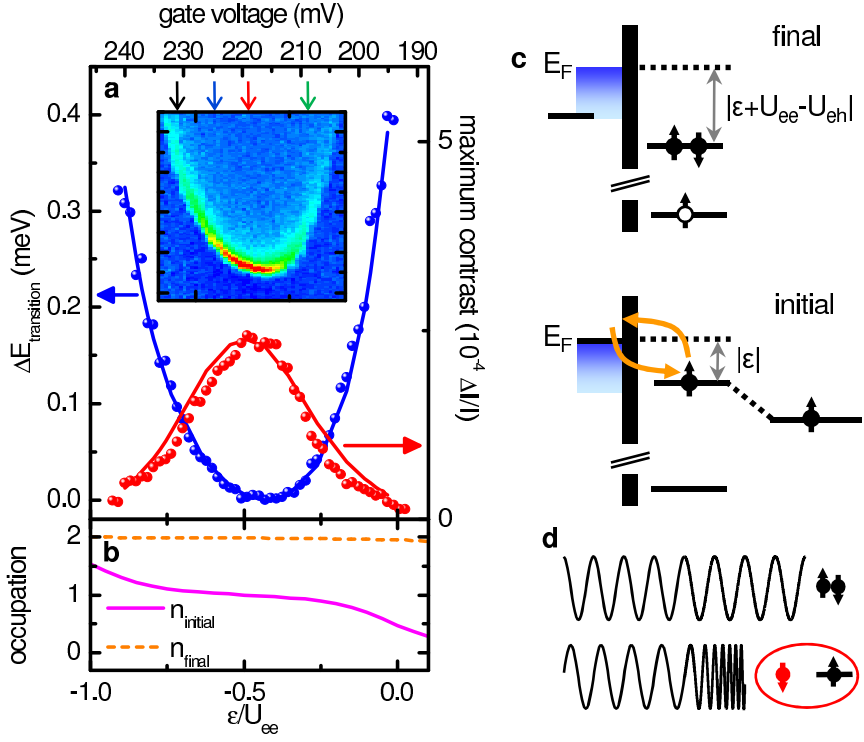}
 \caption{ The gate voltage dependence of the peak absorption
 strength of the negatively charged exciton $\rm{X}^-$, measured
at $180$~mK.
 (a) Inset: absorption as a function of the gate voltage.
 Main figure: Experimental data (symbols) for the
$\varepsilon$-dependence of the shift in the resonance energy
$\Delta E_{\rm transition}$ (blue, left axis) and the absorption
contrast (red, right axis) are well reproduced by NRG calculations
(solid lines) for the following parameters: $\Uee = 7.5$~meV,
$\Gamma = 0.7$~meV, $D = 3.5$~meV, $\Ueh = 11$~meV, $\TFR =
180$~mK.
    (b) Lower panel: NRG results for the occupancy of the QD
   electron level in the initial and final ground states.
    (c) Schematic
  of the energy renormalization process: The initial configuration (bottom)
  features a single
  electron in the QD, whose energy is lowered by virtual tunneling
  between QD and FR.  Since virtual excitations with energy $\Delta E$
  contribute a shift proportional to $- \Gamma/ \Delta E$, the total
  shift (involving a sum over all possible $\Delta E$), is strongest
  near the edges of the X$^-$ plateau. Toward the right edge
  ($\varepsilon$ near 0), the dominant contribution comes from virtual
  tunneling of the QD electron into the FR (as depicted); toward the
  left edge ($\varepsilon$ near $-\Uee$), it comes from virtual
  tunneling of a FR electron into the QD (not depicted).  In the final
  configuration (top), the QD contains two electrons and a hole. The
  electron-hole Coulomb attraction $\Ueh$ effectively lowers the QD
  electron level energy to $\varepsilon - \Ueh$. This raises the
  energy costs $\Delta E$ for virtual excitations by $\Ueh - \Uee$
  (which is $\gg \Gamma$), so that final state energy renormalization
  is negligible.  The renormalization of the transition energy (red
  arrow), probed by a weak laser, is thus mainly due to initial state
  energy renormalization.
  (d) Cartoon for Anderson
    orthogonality: the Kondo cloud (bottom) and local singlet (top) of
    the initial and final configurations produce strong or weak phase
    shifts, respectively.}
\end{figure}

\clearpage
\begin{figure}[t]
    \includegraphics[scale=1]{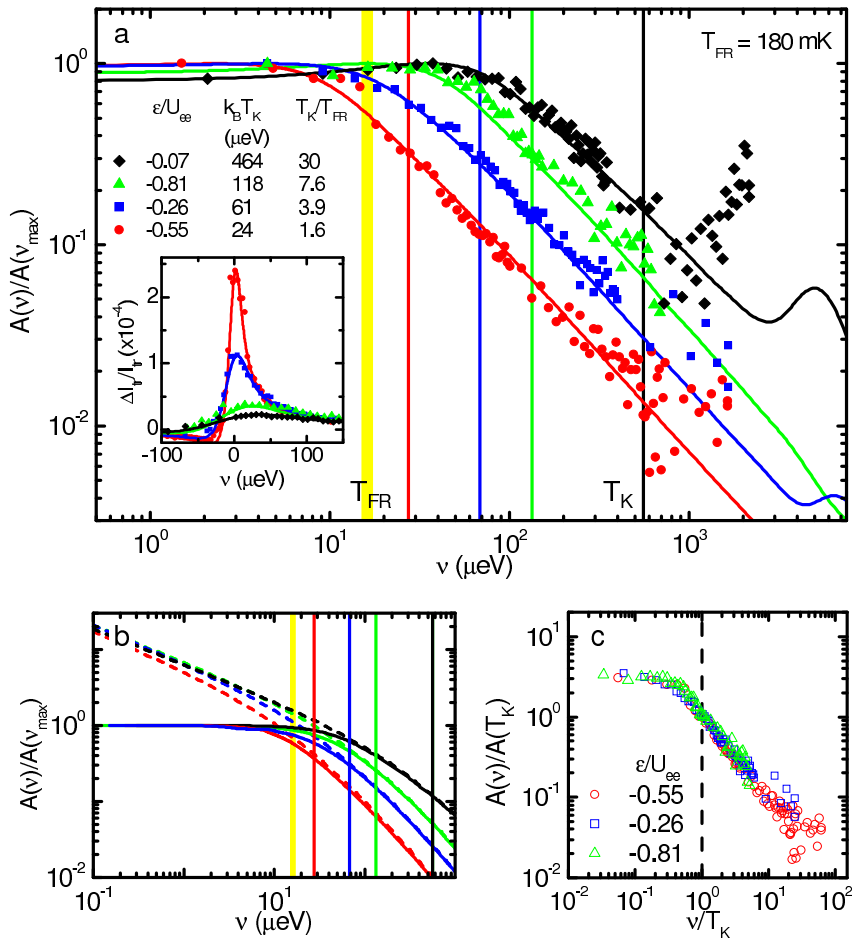}
  \caption{The absorption line shape $A(\nu)$.
  \textbf{a} The blue tail of
    $A(\nu)/A(\nu_{\rm max})$, plotted versus the laser detuning $\nu$ on a log-log
    scale.  The experimental data were taken at an electron temperature of $\TFR =
      180$~mK  for the four values of gate voltage ($\varepsilon$) indicated by arrows in
      Fig.~2\textbf{a}; the corresponding Kondo temperatures $\Tk
      (\varepsilon)$ are indicated by vertical lines in matching
      colors.  NRG results (solid lines), obtained using the
    parameters from the fit in Fig.~2\textbf{a}, are in remarkable
    agreement with experiment. Inset: the measured full (unnormalized) absorption line shape in linear scale
    for the identical $\varepsilon$ values.
    \textbf{b}, NRG results for $T=\TFR$ and $T=0$;
    the latter show the $\nu^{-0.5}$ behaviour expected in the
    strong-coupling regime, $T \ll \nu \ll \Tk$. \textbf{c}, The
    rescaled lineshape $A(\nu)/A(\nu_{\rm max})$ versus $\nu/\Tk$
    shows a universal scaling collapse characteristic of Kondo
    physics.}
\end{figure}

\clearpage
\begin{figure}[t]
   \includegraphics[scale=1]{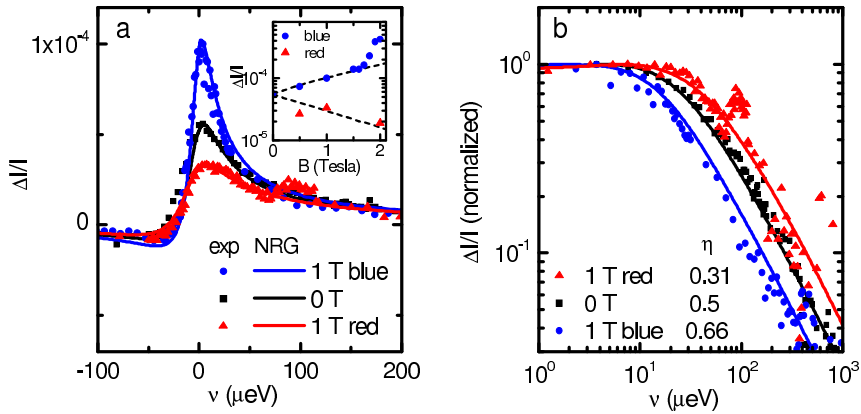}
    \caption{Magnetic field dependence of the absorption.
    (a) The absorption line
    shapes for a second QD with similar parameters (see text)
    for $\varepsilon=-0.43$ at $B_{ext} = 0$~Tesla and
    $B_{ext} = 1$~Tesla for the blue/red trion transition.
     The magnetic field changes the strength of the AOC and the
     lineshape. Inset: the peak absorption contrast showing good
     agreement with the NRG calculations for $B \le 1.5$Tesla.
     (b) The normalized absorption line shape in a log-log plot.
     These measurements pin the value of $\eta(B_{ext}=0)$ to $\sim 0.5$ -
     a direct signature of a Kondo singlet in the absorption lineshape.
     In addition they demonstrate the tunability of an orthogonality
     exponent.}
\end{figure}

\end{document}